# An elaborated pattern-based method of identifying data oscillations from mobile device location data

**Qianqian Sun** (qsun12@umd.edu)
Graduate Research Assistant
Department of Civil and Environmental Engineering
University of Maryland, 1173 Glenn Martin Hall, College Park, MD 20742, United States
ORCID: 0000-0003-3684-4603

**Aref Darzi** (adarzi@umd.edu)
Faculty Assistant
Department of Civil and Environmental Engineering
University of Maryland, 1173 Glenn Martin Hall, College Park, MD 20742, United States

**Yixuan Pan** (ypan1003@umd.edu)
**Corresponding Author**
Research Scientist
Department of Civil and Environmental Engineering
University of Maryland, 1173 Glenn Martin Hall, College Park, MD 20742, United States

**ABSTRACT**
In recent years, passively collected GPS data have been popularly applied in various transportation studies, such as highway performance monitoring, travel behavior analysis, and travel demand estimation. Despite multiple advantages, one of the issues is data oscillations (aka outliers or data jumps), which are unneglectable since they may distort mobility patterns and lead to wrongly or biased conclusions. For transportation studies driven by GPS data, assuring the data quality by removing noises caused by data oscillations is undoubtedly important. Most GPS-based studies simply remove oscillations by checking the high speed. However, this method can mistakenly identify normal points as oscillations. Some other studies specifically discuss the removal of outliers in GPS data, but they all have limitations and do not fit passively collected GPS data. Many studies are well developed for addressing the ping-pong phenomenon in cellular data, or cellular tower data, but the oscillations in passively collected GPS data are very different for having much more various and complicated patterns and being more uncertain. Current methods are insufficient and inapplicable to passively collected GPS data. This paper aims to address the oscillated points in passively collected GPS data. A set of heuristics are proposed by identifying the abnormal movement patterns of oscillations. The proposed heuristics well fit the features of passively collected GPS data and are adaptable to studies of different scales, which are also computationally cost-effective in comparison to current methods.

# 1. Introduction

Large stream GPS data (in the format of device identifier, timestamp, latitude and longitude) is getting popular in many fields. Especially in transportation field, GPS data, as a representation of real-world mobility, have been showing its value in supporting mobility-related studies such as highway performance evaluation, travel statistics estimation, and travel demand estimation. In cases when traditional survey is unavailable or unfeasible because of high time or financial cost, GPS data serve as an alternative data basic. As for GPS-data-driven studies, careful data preparation is essential and critical to ensure reliable mobility inferences. Otherwise, the inferences may be undermined or biased [1, 2, 3]. A major data preparation work that GPS data needs is the detection and removal of data oscillations, aka outliers or data jumps. Although data oscillations (interchangeably used with outliers) are mostly discussed in the context of cellular data or cellular tower data, they also exist in GPS data. However, current methods for detecting outliers in GPS data are either unreliable or focusing on abnormal trajectories instead of abnormal points, and methods developed for cellular data are not applicable to GPS data. This paper aims to identify data oscillations, with points instead of trajectories, as the basic unit, in passively collected GPS data. Passively collected GPS data are particularly discussed since they are more widely applied in practice. Passively collected GPS data have two major sources: direct collection from Global Positioning System (GPS) and indirect collection from location-based services (LBS). Since LBS data relies on the triggering of location service from users or platforms, passively collected GPS data may show differing data features (e.g. frequency, spatial density) at different time windows or locations. This diversity adds challenges to differentiating oscillations from normal points. Most outlier detection methods are statistical tests based on the statistical distribution, e.g. normal distribution, gamma distribution, of a single variable [4, 5, 6]. This method does not apply to passively collected GPS data, which usually does not follow a specific distribution. Current outlier detection methods for trajectory data or GPS data can be classified into speed-based method, distance-based method, partition-and-detect method, density-based method, clustering- or classification-based method, and pattern-based method.

   Speed-based method simply removes GPS points with an extreme high speed from its previous point. Several different speed thresholds are used in studies, such as 200 km/hr [7] and 200 mph [8]. This type of method is unreliable. In reality, the cases of data oscillations in passively collected GPS data are much more complicated. A single variable - speed - is not sufficient for labeling oscillations and may even misidentify normal points as oscillations. For example, the point with an extremely high speed from its previous point may actually be a normal point while its previous point is the true data oscillation. This method may simply remove the later point and keep removing several following points too. Many false positive cases may be caused by this method. A revised speed-based method [9] is later on proposed. Still based on the idea of unreasonable high speed (155 mph), the authors propose a recursive look-ahead filter and Kalman filter to remove outliers. The recursive look-ahead filter firstly presents the value of recursively checking and secondly, when a pair of points show abnormally high intermediate speed, decides which one is the outlier by checking the distance of the two points from the third following point. These two ideas provides reference value for this paper. Kalman filter assumes that the GPS points are vehicle trajectories from a linear dynamical system, whose application is limited since in practice the GPS data are not just vehicle-specific. This paper intends to propose a general method that is not limited to a specific travel mode.

   Distance-based outlier detection method [6] is an early study that detects outliers from trajectory data. It first characterizes each trajectory by a multi-dimension vector with a set of mobility features (i.e. start and end point, number of points, heading, and velocity). Then a distance function is built by the weighted sum of the difference between the vectors of two trajectories to measure the similarities between paths. This method has some limitations. If there is just a small segment showing abnormal features in a very long trajectory, this abnormality is probably averaged out across the entire trajectory so that cannot be identified [10]. To make outliers stand out, most of the trajectories in a dataset have to share similar mobility behavior. However, in many practical cases, GPS trajectory data show various mobility behavior even if they are not outliers. Additionally, this method removes the entire abnormal trajectory, which results in unnecessary data loss.

Instead of identifying and removing abnormal trajectories, some studies focus on the outlier segments in a trajectory. These studies apply a partition-and-detect model [10-15]. The partition-and-detect model first partitions trajectories and then removes abnormal sub-trajectories. Abnormal sub-trajectories are detected from aspects such as position and angle by taking neighboring trajectories as references. A major limitation of this kind of method is that detecting outliers depends on neighboring trajectories. With the absence of neighboring trajectories, the standalone sub-trajectories are determined to be outliers and therefore removed. This method necessitates a high spatial density of neighboring trajectories; otherwise, normal sub-trajectories are misidentified as outliers and data loss is too high in this case.

There are also density-based outlier detection methods, which are firstly proposed by defining local outlier factor (LOF) [16]. LOF measures how a point is isolated from its neighboring points. Points with a high LOF are determined to be outliers. Neighborhood size plays a critical role for this method since if it is too small outliers may not be identified and if it is too large too much computation cost is required. The neighborhood size proves to be very sensitive [17]. The LOF involves too much computation between neighboring points, which is not a good choice for large-scale (e.g. statewide or nationwide) datasets. Passively collected GPS data may have low frequency for a while making it normal to have standalone points that are geographically distant from others. They are not outliers and may be an intermediate point on a trip. However, density-based methods may mistakenly remove them.

Similar to density-based methods, some clustering-based methods [18, 19] detect outliers based on spatial densities. However, clustering-based methods originally intend to generate clusters instead of to identifying outliers. The points that cannot form or be involved in a cluster are identified as outliers, which may actually be normal points in other research contexts. Passively collected GPS data may present different spatial densities at different locations making it tricky to define cluster sizes. Additionally, low-density points that are unable to form a cluster are not necessarily outliers and instead may be essential for other contexts such as threshold-based trip identification studies. Classification-based method [20] is proposed for detecting abnormal trajectories by leveraging a good training dataset, which is unavailable in passively collected GPS data.

One more type of method for detecting data oscillations is pattern-based methods. It is originally proposed [21] for identifying the ping-pong transitions in cellular data (the cell that a mobile connects to changes in a short time making its geographical location changes quickly back and forth). This method defines two movement patterns of data oscillations by the sequence of cellular towers. Some later studies made improvements on this method to identify the ping-pong phenomenon in different ways, such as additionally considering spatial-temporal features [7], building clusters [3], and defining different time windows [2]. All these methods are specifically designed for the ping-pong phenomenon in cellular tower data or cellular data. GPS data have a higher location precision and accuracy than cellular data, so GPS movement is more granular and uncertain making it difficult to define abnormal movement patterns. Passively collected GPS data usually present varying sampling frequencies making the data oscillations show much more various and complicated patterns than the ping-pong oscillations. Although the pattern-based methods are insufficient and inapplicable to be directly applied to passively collected GPS data, the authors see the merits of pattern-based methods. In this paper, several heuristics are proposed to address oscillations in passively collected GPS data with revisions and improvements on a state-of-the-art pattern-based method.

## 2. Data Oscillation Identification

As afore-discussed, pattern-based methods show the greatest reference value for developing methods for identifying data oscillations (i.e. abnormal points) from passively collected GPS data. Although they are specifically designed for cellular data's ping-pong phenomenon, the core idea is enlightening: data oscillations usually present abnormal movement patterns and impossibly-high moving speeds [1 - 3, 7, 21]. State-of-the-art pattern-based method [1] designs several heuristics to identify data oscillations from cellular data, which is revised and improved for the application to the passively collected GPS data. Additionally, the ideas of being recursive and additionally checking the third following point from the recursive look-ahead filter [9] as previously discussed are incorporated in this paper. In this study, the revisions and improvements are made from five aspects.
- First, a flexible spatial-temporal formulation is constructed by using adaptive and dynamic parameters instead of static parameters, which includes more cases of data oscillations.
- Second, location uncertainty is reduced by using geohash zones instead of raw latitude and longitude or building clusters, which saves much computation cost and is much more cost-effective in large-scale studies.
- Third, inspired by "stable period", the definition of "stable community" and "stable zone" is designed to label the GPS points that are believed to be normal points, which is more flexible by building local communities and is more reliable by additionally consider dwell time and frequency.
- Forth, more movement patterns of data oscillations are included by additionally considering data oscillations that continuously happen one by one.
- Fifth, parameters are chosen by sensitivity tests or empirical experience from literature review instead of being arbitrarily decided.

In order to denote movement patterns, studies usually transform the raw latitude and longitude location into clusters [3]. To reduce computation cost, another method is applied: latitude and longitude coordinates are projected to level-7 geohash zones (152.9 meters × 152.4 meters near the equator). This greatly reduces the location uncertainty of GPS points and serves as the basis of formulating movement patterns. Hence, the raw GPS movement of each device is denoted by a sequence of level-7 geohash zones. Abnormal movement patterns are proposed and data oscillations are removed accordingly.

## 2.1. Stable Communities and Stable Zones

A major assumption is proposed that if a device is frequently observed (multiple GPS sightings) or is observed long enough (long dwell time) at a location, this location is believed to be a true visit. The true visit of a location here does not refer to a level-7 geohash zone. Instead, a dynamic community is built to represent a location since devices are not just staying static at a place and in most cases they are moving on the way. The community here is a local group of continuous sightings for a given device during a time period. Several different communities can form for a device. Each community grows by including more and more continuous sightings that are close enough, i.e. less than $dist_c$ miles between two continuous sightings. Since the traces of a device are represented by a sequence of level-7 geohash zones, a community built upon this is also a group of level-7 geohash zones. Sometimes, a community can only contain one single geohash zone including one or multiple GPS points. $dist_c$ is determined based on the statistical analysis of the GPS data. For example, a tested GPS dataset shows that the distance between two continuous GPS sightings for each device has a mean value of 0.4 miles. So its rounded value of 0.5 miles can be the threshold value to construct a community with a tolerance of the marginal variations.

Note that a community could be a group of data oscillations. Based on the investigation of testing data, when one data oscillation occurs, additional oscillations could happen surrounding the first oscillation. For the purpose of differentiating true communities and oscillation communities, the definition of the stable community is proposed, which means if a device presents enough occurrences or dwells long enough in a community, this community is believed to be a true visit and is determined to be a stable community. All geohash zones included by a stable community are defined as stable zones and all GPS points in a stable community are determined to be true normal points. Each community has two attributes: frequency and duration. Frequency is the total number of GPS points and duration is the time interval between the first and the last point. The threshold of frequency can be determined based on the analysis of the GPS points. For example, testing data show that the mean value of number of GPS points in each level-7 geohash zone is five. The threshold of duration can be 300 seconds since it is a state-of-the-practice dwell time value used for determining trip stops from GPS data. A sensitivity analysis of the testing data (Figure 1) also shows that the number of identified oscillations reaches the peak of 1.48 when the duration threshold value is set as 300 seconds.

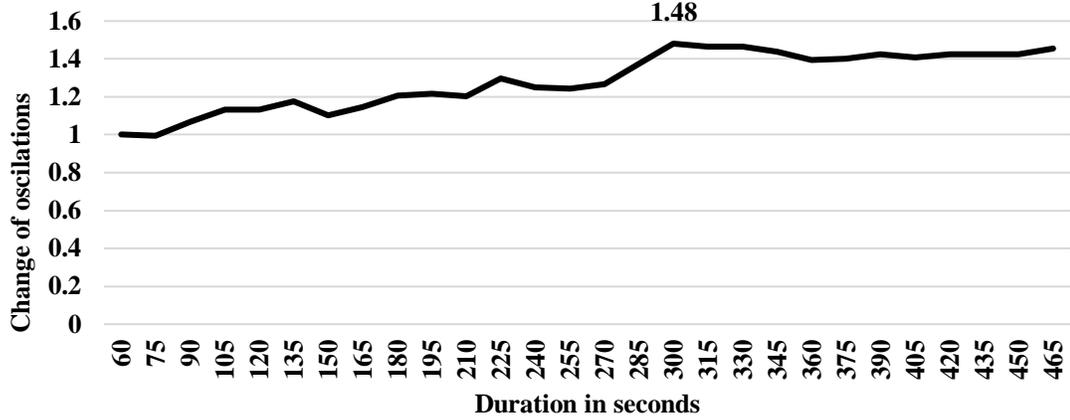

Figure 1. Sensitivity Analysis of the Duration Threshold

Figure 2 denotes three communities of a device, among which two are stable and one is unstable. The blue community consists of two level-7 geohash zones and the green community consists of three level-7 geohash zones. They are stable communities, i.e. assumed to be true visits, since the data frequency is high (e.g. more than five GPS points). The red community is unstable, which only contains one GPS point. Being unstable does not mean being oscillations. They simply lose the candidacy of being referenced as true normal visits when detecting nearby oscillations.

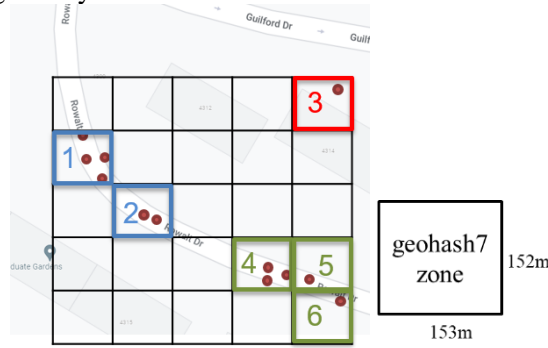

Figure 2. Example of Stable and Unstable Communities

### 2.2. Heuristics for Identifying Oscillations

Based on the definition of stable communities and stable zones, two heuristics at level-7 geohash zone level (Heuristic 1a, 1b) and two heuristics at the community level (Heuristic 2a, 2b) are developed to identify data oscillations. They are designed to be recursively applied to the GPS data. The details are described below.

- Heuristic 1a: it is assumed that a device cannot finish a round trip within an extremely short time interval and a device cannot move to a faraway place at an abnormally high speed using ground transportation. For example, if a device, within 30 seconds, starts from a stable level-7 geohash zone and goes back to the same level-7 geohash zone that is also in stable status, all the middle sightings are determined as data oscillations and removed. The time threshold $t_{min}$ is calculated by the equation: $t_{min} = 2 * dist_c/v_{max}$. When $dist_c = 0.5$ miles and $v_{max} = 155/1.3 = 120\ mph$, $t_{min} = 30$ seconds. $dist_c$ is used since when the middle geohash7 zone(s) are unstable zones, it has to be at least $dist_c$ far away. Otherwise, it is in the same stable community as its previous community. 155 miles per hour is twice the allowed driving speed [9]. Since geodesic distance instead of road network distance is used in the equation, a detour factor of 1.3 is used for making up for the difference. Figure 3 is an example of heuristic 1a, the device starts from a stable level-7 geohash zone on the top, then jumps to another zone on the right below, then jumps back to the same zone as the beginning. The middle point on the right is determined to be the data oscillation.

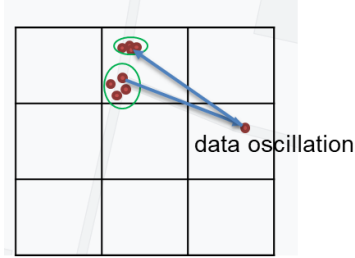

Figure 3. Example of Heuristic 1a

- Heuristic 1b: considering that passively collected GPS data have varying data frequency and may be low frequent sometimes, it may be too demanding to satisfy the detection criteria mentioned above. Thus, a supplementary check is conducted. It is assumed that a device cannot move to a faraway place at an abnormally high speed using ground transportation. For any pair of two consecutive level-7 geohash zones, if one is stable while the other is unstable, the unstable zone is investigated to see if any oscillation exists. If the distance between the two level-7 geohash zones is more than $dist_g$ (e.g. 5 miles) and the time interval between the two zones is less than $t_g$ (e.g. 2.5 minutes), then all sightings in the unstable zone are determined to be data oscillations and removed. Instead of a single speed parameter, distance and time interval are additionally considered. This is because that drifting GPS points are common which may show very high speed but are actually not the true data oscillations. These drifting points do not have bad effects on the inferences as data oscillations. This heuristic is exemplified for cellular data [1] using 5 km and 1 minute. Here, the heuristic is revised by applying level-7 geohash zones and using an adjusted distance threshold of 5 miles and time threshold of 2.5 minutes, whose corresponding speed is 120 mph (the maximum allowed speed with detour factor considered). The thresholds are subject to change on a case-by-case basis. Figure 4 is an example of heuristic 1b. A device forms a stable level-7 geohash zone (sg2), its previous zone (g1) and its following zone (g3) are checked for possibly being outliers. When the two thresholds ($dist_g$, $t_g$) are satisfied, data oscillations are identified such as the point included in g3.

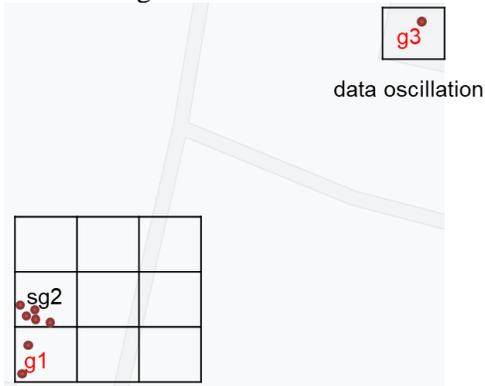

Figure 4. Example of Heuristic 1b

- Heuristic 2a: the oscillations considered here are these obvious ones from geometry point of view. Any sequence of three consecutive communities of a device can form a triangle. The distance and speed between communities are checked against two conditions as below. When the two conditions are satisfied, all GPS points in the middle community are probably oscillations and are removed.

$$v_{i-1,i} * v_{i,i+1} > 155 * 155$$
$$dist_{i-1,i+1} < 0.25 * \min(dist_{i-1,i}, dist_{i,i+1})$$

where $v_{i-1,i}$ and $dist_{i-1,i}$ are the speed and the distance respectively from the first community to the second community; similarly, $v_{i,i+1}$ and $dist_{i,i+1}$ are the speed and the distance from the second community to the third community; $dist_{i-1,i+1}$ is the distance between the first community and the third community. As shown in the Figure 5, if the two conditions are satisfied, the points included in the middle community (C2) are determined to be data oscillations.

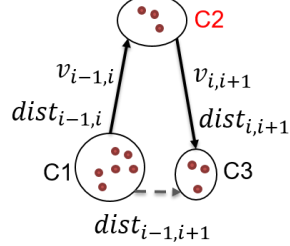

Figure 5. Example of Heuristic 2a

- Heuristic 2b: sometimes, the case described by heuristic 2a can continuously occur, i.e. a set of communities with every three continuous ones satisfying the two conditions in heuristic 2a. In this case, true oscillations need to be decided with additional consideration. Starting from the first community, all communities are labeled with a position number in numerous order (i.e. 1, 2, 3, …). Therefore, there are two groups of communities: odd and even position communities. It is assumed that communities with longer average dwell time are the true normal visits and are kept. As shown in Figure 6, it is continuously observed that a trio of communities, e.g. (C1, C2, C3), (C2, C3, C4), ……, (C6, C7, C8), meets the two conditions in heuristic 2a. Then the group of either even position or odd position is kept based on the average dwell time across communities in each group.

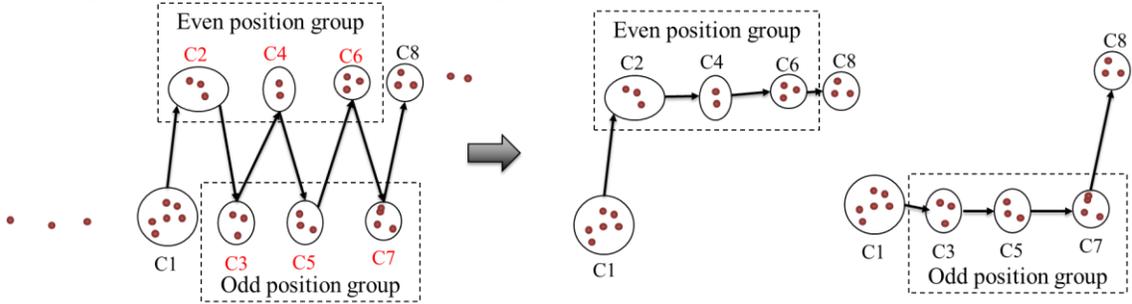

Figure 6. Example of Heuristic 2b

## 3. Discussion

GPS data is widely applied in various transportation studies while data preprocessing especially removing oscillated points is overlooked, which initiates this paper. The authors conduct a thorough review of previous studies on data oscillations from location data (i.e. GPS point and trajectory data, cellular data, and cellular tower data) and discuss the advantages, shortcomings, and applicability of passively collected GPS data. Considering the features of oscillations in passively collected GPS data (i.e. higher uncertainty, diversity, and complexity), some merits are adopted from and several revisions and improvements are made to previous studies. To the best knowledge of the authors, the way of reducing location uncertainty and defining movement patterns through geohash zones is for the first time discussed. In comparison to other methods of using raw latitude and longitude or clusters, this method is not only computationally cost-saving but also shows high flexibility and adaptability to various case studies since the level of geohash zoning can be adjusted. For high-frequency GPS data, a higher level geohash zone system (e.g. level-8 geohash zone with a size of 38.2 meters x 19 meters near the equator) can be adopted. Additionally, various patterns of oscillations in passively collected GPS data are incorporated through two levels of heuristics based on the innovative idea of stable communities and stable geohash zones. Instead of removing oscillated location points in GPS data simply based on speed like many current studies do, this paper firstly makes assumptions about the normal points and then checks other points for being oscillations or not based on their spatial-temporal relationship from the normal points. Considering the different scales that an oscillation may show in passively collected GPS data, dynamic parameters are used. Although there have been well-developed data oscillation identification methods, these methods are designed for cellular data, which are unable to handle oscillations in passively collected GPS data. Other GPS-related oscillation identification methods all have different shortcomings: speed-based may misidentify normal points as outliers; density-based and clustering-based methods may misidentify standalone points as outliers, which may commonly exist in passively collected GPS data; distance-based and partition-and-detect method conditions on high spatial density dataset and can cause high data loss when applied to low-density datasets. The heuristics in this paper are designed in a way of carefully identifying oscillations and are applicable to passively collected GPS data with varying data densities at different locations, which are of practical significance.

**ACKNOWLEDGMENTS**
We would like to thank and acknowledge our partners and data sources in this effort: (1) partial financial support from the U.S. Department of Transportation's Federal Highway Administration; (2) Amazon Web Services and its Senior Solutions Architects, Jianjun Xu and Greg Grieff, for providing cloud computing and technical support.

**AUTHOR CONTRIBUTIONS**
The authors confirm their contribution to the paper as follows: study conception and design: Q. Sun, Y. Pan, A. Darzi; data collection and preparation: Q. Sun; analysis and interpretation of results: Q. Sun, Y. Pan, A. Darzi; draft manuscript preparation: Q. Sun, Y. Pan, A. Darzi. All authors reviewed the results and approved the final version of the manuscript.